\documentclass[a4paper,12pt]{article}
\usepackage{amsmath}
\usepackage{graphicx}

\def\scr{\cal}

\begin{document}

\parindent=0pt
\parskip=10pt plus 4pt minus 2pt

\title{A simple expression for the ADM mass}
\author{%
Leo Brewin\\[10pt]%
School of Mathematical Sciences\\%
Monash University\\%
Clayton 3800\\%
Australia}

\maketitle

\begin{abstract}
We show by an almost elementary calculation that the ADM mass of an
asymptotically flat space can be computed as a limit involving a rate of change
of area of a closed 2-surface. The result is essentially the same as that given
by Brown and York \cite{brown-york:1993-01,brown-lau-york:1997-01}. We will
prove this result in two ways, first by direct calculation from the original
formula as given by Arnowitt, Deser and Misner and second as a corollary of an
earlier result by Brewin for the case of simplicial spaces.
\end{abstract}

\section{Introduction}

It is well known that the for the Schwarzschild spacetime the ADM and the
Schwarzschild masses are equal. Interestingly one can also easily recover the
Schwarzschild mass by a simple calculation involving the areas of 2-spheres.
Consider the time symmetric initial data for the Schwarzschild spacetime
in the standard Schwarzschild coordinates
$$
ds^2 = \left(1-\frac{2m}{r}\right)^{-1}dr^2 + r^2d\Omega^2
$$
The area of a sphere of radius $r$ is $A = 4\pi r^2$ and its rate of change
with respect to proper radial distance is
$$
\frac{dA}{ds} = 8\pi r\left(1-\frac{2m}{r}\right)^{(1/2)}
$$
For $r>>2m$ we can expand this in powers of $m/r$ to obtain
$$
\frac{dA}{ds} = 8\pi r - 8\pi m + {\cal O}\left(\frac{1}{r}\right)
$$
The leading term on the right hand side is exactly the flat space value for
$dA/ds$ and thus we can re-write the above in the suggestive form
$$
 8\pi m =   \left(\frac{dA}{ds}\right)_f
          - \left(\frac{dA}{ds}\right)_c
          + {\cal O}\left(\frac{1}{r}\right)
$$
with the subscripts `f' and `c' denoting flat and curved space quantities 
respectively. The main result of this paper is that a similar result applies to
all asymptotically flat spacetimes.

The first point to address is how we might compute the curved and flat space
versions of $dA/ds$ for a general 3-metric. The procedure we shall follow is
quite simple to state. We start with a given space $({\cal M},g)$ and then chose
a large closed 2-surface $S$ in ${\cal M}$.  For later convenience we shall
assume that the surface $S$ is everywhere convex (i.e. that its intrinsic
Gauss curvature is everywhere non-negative). This is not a serious constraint
and should be easily satisfied provided the surface is built sufficiently far
into the asymptotically flat regime of the space. 

The surface $S$ will inherit, from its embedding in $({\scr M},g)$, a unit
normal $n^\mu$ and an intrinsic 2-metric $h_{\mu\nu}$. Thus we can compute not
only the surface area $A$ of $S$ but also the surface areas for images of $S$
obtained by dragging $S$ a short distance along the integral curves of $n^\mu$.
This in turn will allow us to compute the rate of change of the area, $dA/ds$,
with respect to the proper distance $s$ off $S$.

Now we shall engage in a small piece of surgery. First we discard the portion of
$({\scr M},g)$ outside of $S$ (though we retain $S$ itself). Since, by
assumption, $S$ is everywhere convex we are assured by the Weyl-Lewy embedding
theorem \cite{lewy:1938-01,nirenberg:1953-01} that the truncated space may be extended by
attaching a sub-set of flat space to $S$. The result is a new space which we
denote by $(\tilde{\scr M},\tilde g,S)$. The surface $S$ can now be viewed as a
transition surface from the curved interior to a flat exterior. We shall use a
{`$+$'} subscript to denote quantities on the exterior and likewise a {`$-$'}
subscript for quantities associated with the interior. Note that the surface
components of the metric (i.e. the projection of $\tilde g$ onto $S$) will be at
least $C^0$ but no such continuity can be expected for the normal components
(i.e. the projection onto the normal to $S$). Thus we can not expect the area $A$
to vary smoothly as we cross $S$. Indeed we now claim that, in the limit
$S\rightarrow i^0$, the jump in $dA/ds$ is directly proportional to the ADM
mass,
\begin{equation}
M = \frac{1}{8\pi} \lim_{S\rightarrow i^0}
       \left( \left(\frac{dA}{ds}\right)_{\tt +}
             -\left(\frac{dA}{ds}\right)_{\tt -} \right)
\label{ADMmassA}
\end{equation}

We will present two very simple yet independent proofs. The first
applies to smooth metrics and begins with the standard expression for the ADM mass
in asymptotic Euclidian coordinates \cite{adm:1962-01,mtw:1973,wald:1984} while the
second proof, applicable to simplicial spaces, follows as a simple corollary to an
earlier result by Brewin \cite{brewin:1989-01}.

Brown and York \cite{brown-york:1993-01,brown-lau-york:1997-01} have produced
a very similar result (with a slight change of notation),
\begin{equation}
M = \frac{1}{8\pi}\lim_{R\rightarrow\infty}\>\int\left(K_{-} - K_{+}\right)\>dS
\label{BrownYork}
\end{equation}
where $K_{+}$ and $K_{-}$ are the traces of the intrinsic curvatures of $S_{+}$
and $S_{-}$ respectively. By noting that for any 2-surface embedded in a 
larger ambient 3-space, $dA/ds= -\int K dS$ it follows that two expressions
(\ref{ADMmassA},\ref{BrownYork}) for $M$ are equivalent.

What then is new in the present calculation? Simply that we offer an
alternative, and perhaps simpler proof of the main result (\ref{ADMmassA}). This
fits the maxim that simple results demand simple proofs. 

Our calculations are adapted to just one quantity, the ADM mass. In contrast
Brown and York's computations are much more extensive. They derive quasi-local
estimates for the energy and momentum and they show that these reduce to the
familiar ADM expressions at spacelike infinity. Their calculations are much more
sophisticated than ours. In our approach we will compute the ADM mass directly
from the standard integral formula.

\section{The ADM mass}

Given a 3-manifold ${\scr M}$ with an asymptotically flat positive definite
3-metric $g$ the ADM mass of the space $({\scr M},g)$ can be defined by
(\cite{adm:1962-01,mtw:1973,wald:1984})
\begin{equation}
M = \frac{1}{16\pi}\lim_{S\rightarrow i^0}\int_S\>
g^{\mu\nu}\left(g_{\mu\alpha,\nu} - g_{\mu\nu,\alpha}\right)n^\alpha\> dS
\label{ADMmassB}
\end{equation}
where $S$ is a topological 2-sphere with outward pointing unit normal $n^\mu$
and area element $dS$. Note that the integrand is not in covariant from (indeed
there is no covariant expression involving the first derivatives of the metric
other than $g_{\mu\nu;\alpha}=0$). Thus the above integral is defined only for a
restricted class of coordinate systems and these are known as asymptotic
Euclidian coordinates. In such a system the metric components are required to
be of the form $g_{\mu\nu} = \delta_{\mu\nu} + {\cal O}(1/r)$. 

We first prepare for a series of local calculations by  sub-dividing $S$ into a
finite set of non-overlapping patches $S_i,i=1,2,\cdots$. We do this simply to
avoid any complications that might arise from doing a single global calculation.
For much of the following we will focus on a single patch which we shall denote
by $S_i$. Our next step will be to show that a simple coordinate system can be
constructed in a subset of ${\cal M}$ containing $S_i$. The construction is
quite elementary. Consider a typical $S_i$ of $S$ and build a short cylinder
$C_i$, with cross-section $S_i$, by following the integral curves of the normal
vectors to $S$. The cylinder will extend partly into ${\scr M}_{+}$ and partly
into ${\scr M}_{-}$. On the patch $S_i$ construct a 2-dimensional set of 
coordinates $(u,v)$. If the cylinder is sufficiently short then each point $P$
in the cylinder will be threaded by exactly one of the integral curves
originating from $S_i$. This allows us to assign coordinates $(u,v,n)$ to $P$ in
the obvious way, namely $n$ is the (signed) proper distance to $P$ measured
along the integral curve from $S_i$ and $(u,v)$ are the coordinates in $S_i$
where the integral curve intersects $S_i$. We take $n>0$ for points in ${\scr
M}_{+}$ and $n<0$ for points in ${\scr M}_{-}$. We will now make a coordinate
transformation in ${\scr M}_{+}$. Since ${\tilde g}_{+}$ is flat we know that we
can transform the coordinates in ${\scr M}_{+}$ such that ${\tilde g}_{+} =
diag(1,1,1)$. Finally we choose any smooth extension of the coordinate
transformation into ${\scr M}_{-}$. By this construction we will have built a
(locally) asymptotically Euclidian coordinates in the cylinder $C_i$ thus allowing
us to use the above integral (\ref{ADMmassB}) for the contribution from
$S_i$ to the ADM mass $M$.

Let $A_i$ be the surface area of $S_i$. Then using standard results from
differential geometry we have
$$
\frac{dA_i}{ds} = \frac{1}{2} \int_{S_i}\>
   g^{\mu\nu}{\cal L}_n\>g_{\mu\nu}\>dS
$$
where $dS$ is the area element on $S_i$ and $n^\mu$ is the unit normal vector to
$S_i$. However
\begin{eqnarray*}
g^{\mu\nu}{\cal L}_n\>g_{\mu\nu}
  &=& 
   g^{\mu\nu}\left(
       g_{\mu\nu,\alpha}n^\alpha
     + g_{\alpha\nu} n^\alpha{}_{,\mu}
     + g_{\mu\alpha} n^\alpha{}_{,\nu}\right)\\[5pt]
   &=& 
   g^{\mu\nu}g_{\mu\nu,\alpha} n^\alpha
     + g^{\mu\nu}\left( g_{\alpha\nu} n^\alpha\right)_{,\mu}
     - g^{\mu\nu}g_{\alpha\nu,\mu} n^\alpha
     + g^{\mu\nu}g_{\mu\alpha}n^\alpha{}_{,\nu}\\[5pt]
   &=& 
   g^{\mu\nu}\left( g_{\mu\nu,\alpha} - g_{\alpha\nu,\mu} \right)n^\alpha
     + g^{\mu\nu} n_{\mu,\nu} + n^\mu{}_{,\mu}
\end{eqnarray*}
and thus we have
$$
\frac{1}{8\pi}\frac{dA_i}{ds} = 
    \frac{1}{16\pi}\int_{S_i}\>
       g^{\mu\nu}\left(g_{\mu\nu,\alpha} - g_{\alpha\nu,\mu}\right)n^\alpha\>dS
   +\frac{1}{16\pi}\int_{S_i}\>\left( g^{\mu\nu} n_{\mu,\nu} + n^\mu{}_{,\mu} \right)\>dS
$$
The integrand has been deliberately re-arranged to isolate the terms that arise
in the original definition of the ADM mass. 

Let us now define $[Q]$ for any object $Q$ to be the jump in $Q$ across $S$,
that is
$$
[Q] = Q_{+} - Q_{-}
$$
Now consider the pair of integrals
\begin{eqnarray*}
I_1 &=& \int_{S_i}\>\big[\mathstrut 
        g^{\mu\nu}\left(g_{\mu\nu,\alpha} - g_{\alpha\nu,\mu}\right)n^\alpha\big]\>dS \\[5pt]
I_2 &=& \int_{S_i}\>\big[\mathstrut 
        g^{\mu\nu} n_{\mu,\nu} + n^\mu{}_{,\mu}\big]\>dS
\end{eqnarray*}
It is not hard to see that our claim (equation \ref{ADMmassA}) is true provided
we can show that the second integral $I_2$ vanishes in the limit $S\rightarrow
i^0$. To prove this we need to carefully examine the asymptotic behaviour of the
integrand. Let us first look at the relationship between $n^\mu{}_{+}$ and
$n^\mu_{-}$. Since $n^\mu_{+}$ and $n^\mu_{-}$ are normal to the surface defined
by $n=0$ we must have $n_{\mu+} = f_{+}n_{,\mu}$ and $n_{\mu-} = f_{-}n_{,\mu}$
where $f_{+},f_{-}$ are scalar functions on $S_i$. Thus we find
$$
n_{\mu-} = f n_{\mu+}
$$
where $f = f_{-}/f_{+}$. We can also establish a similar result for the
components $n^\mu$. To this end, consider any pair of
tangent vectors $a^\mu,b^\mu$ to $S_i$. Clearly $0=a_\mu n^\mu_{+}$ and $0=b_\mu
n^\mu_{+}$ and thus $n^\mu_{+} =
q_{+}\epsilon^{\mu\alpha\beta}_{123}a_{\alpha}b_{\beta}$ where $q_{+}$ is a
function on $S_i$. Likewise, for $n^\mu_{-}$ we find $n^\mu_{-} =
q_{-}\epsilon^{\mu\alpha\beta}_{123}a_{\alpha}b_{\beta}$ and thus $n^\mu_{-}
=(q_{-}/q_{+})n^\mu_{+}$. But since both normals are required to be unit vectors
we must have $q_{-}/q_{+} = 1/f$ and thus
$$
n^{\mu}_{-} = \frac{1}{f} n^{\mu}_{+}
$$
What can we say about the asymptotic behaviour of $f$? We know that in the
conformal analysis of asymptotically flat spacetime the 3-metric is $C^{>0}$ at
$i^0$. The lack of differentiability is intimately tied to the direction
dependent limits at $i^0$. Thus it is not unreasonable to assume that the
function $f$ will vary smoothly along each spacelike geodesic ending at $i^0$.
The limit may well depend on which geodesic is chosen but along any one geodesic
the function $f$ should vary smoothly. Thus we propose that $f$ should posses an
asymptotic expansion of the form
$$
f = 1 + \frac{a}{r} + {\cal O}\left(\frac{1}{r^2}\right)
$$
where $a$ may be a direction dependent quantity near $i^0$. Alternatively one
might argue that the following analysis is applicable only to those spaces for
which $f$ obeys the above form. In either case the class of spaces with this
behaviour admits at least the time symmetric initial data of the Schwarzschild
spacetime.

Now let us turn our attention to the 3-metric and its behaviour across $S$. The
integral curves through $S_i$ provide a natural means for foliating $C_i$ by a
sequence of images of $S_i$. Thus we can write the 3-metric in $C_i$ in a 2+1
form
$$
g_{\mu\nu} = h_{\mu\nu} + n_\mu n_\nu
$$
In this form $h_{\mu\nu}$ is the induced 2-metric on each leaf of the foliation.
Continuity of the 3-metric across $S_i$ requires
$$
0 = [h_{\mu\nu}]\quad\quad {i.e.}\ \ h_{\mu\nu+} = h_{\mu\nu-}
$$
All indices will be raised and lowered using the 3-metric $g_{\mu\nu}$.

We can now express all of the terms in $I_2$ solely in terms of quantities based
in ${\cal M}_{+}$. Thus we obtain
\begin{eqnarray*}
\big[\mathstrut g^{\mu\nu}n_{\mu,\nu} \big]
&=&  \left(h^{\mu\nu} + n^\mu n^\nu\right)n_{\mu,\nu}
    -\left(h^{\mu\nu} + \frac{1}{f^2}n^\mu n^\nu\right)\left(fn_{\mu}\right)_{,\nu}\\[5pt]
&=& (1-f)h^{\mu\nu}n_{\mu,\nu} - \frac{1}{f^2}f_{,\mu}n^\mu\\[5pt]
\big[\mathstrut n^\mu{}_{,\mu}\big ]
&=& n^\mu{}_{,\mu} - \left(\frac{1}{f}n^\mu\right)_{,\mu}\\[5pt]
&=&\left(1 - \frac{1}{f}\right)n^\mu{}_{,\mu} + \frac{1}{f^2}f_{,\mu}n^\mu
\end{eqnarray*}
where all terms on the right hand side are understood to be evaluated in ${\cal
M}_{+}$ and thus, for simplicity, the $+$ subscripts have been dropped. Note that
we have also made use of the fact that the metric in ${\cal M}_{+}$ is flat and
thus in the Euclidian coordinates (which we chose earlier)
$h^{\mu\nu}n_{\mu,\nu} = g^{\mu\nu}n_{\mu,\nu} = n^\mu{}_{,\mu}$ and $0=n^\mu
n^\nu n_{\mu,\nu}$. Combining the above we obtain
$$
 \big[\mathstrut g^{\mu\nu}n_{\mu,\nu} \big]
+\big[\mathstrut n^\mu{}_{,\mu} \big]
= - \frac{1}{f}\left(1-f\right)^2h^{\mu\nu}n_{\mu,\nu}
$$
Notice that the term $h^{\mu\nu}n_{\mu,\nu}$ is proportional to the trace of the
intrinsic curvature on $S_i$ and thus varies as ${\cal O}(1/r)$. Finally, since
$f-1={\cal O}(1/r)$ we see that the integrand of $I_2$ varies as ${\cal
O}(1/r^3)$ and thus $I_2$ will vanish in the limit as $S_i\rightarrow i^0$. This
completes the proof, that $I_2$ vanishes. The last step is to form a sum over
all of the $S_i,i= 1,2,\cdots$ yielding the formula as stated above (equation
\ref{ADMmassA}).

\section{Simplicial lattices}

Lattice theories such as the Regge Calculus (see
\cite{regge:1961-01,williams:1992-01,gentle:2002-01}) have enjoyed some success
as a tool for numerical relativity though they have more commonly been used as a
model for discrete theories of gravity. A popular example is the Regge calculus
in which a simplicial lattice is built from a collection of non-intersecting
blocks (which in 3-d are usually tetrahedral), each with a flat internal
metric, and an assignment of leg lengths to each leg in the lattice.   Though
the metric is flat in each tetrahedron, it is not flat everywhere. In fact the
curvature must be treated as a distribution with support on the legs of the
lattice. This complicates the study of such lattices but progress can be made.
Indeed as an example of the use of standard distribution theory we  showed
\cite{brewin:1989-01} that the ADM mass of a simplicial lattice could be
computed as\footnote{Note that there was a typographical error in Brewin's
paper \cite{brewin:1989-01} . The above corrected formula differ's from Brewin's
by a factor of -1/2}
\begin{equation}
M = - \frac{1}{8\pi} \sum_S\>\theta L
\label{ADMmassC}
\end{equation}
where the sum includes all the legs in the (triangulated) 2-surface $S$,
$\theta$ is the defect angle on the leg and $L$ is the length of that leg. The
defect angle $\theta$ is defined as the $\theta = 2\pi - \sum_i\theta_i$ where
$\theta_i$ are the angles subtended at the leg by the blocks attached to the
leg. In our case we have four triangular cylinders attached to each leg on the
surface $S$, with a pair of cylinders on either side of $S$. The defect angle
can then be computed as $\theta = - \theta_{+} - \theta_{-}$ where $\theta_{+}$
and $\theta_{-}$ are the contributions from the respective pairs of cylinders in
${\cal M}_{+}$ and ${\cal M}_{-}$. Finally, by inspection of Fig. 1, we can see
that $(\theta_{+})L = \left(dA/ds\right)_{+}$ and $(\theta_{-})L =
\left(dA/ds\right)_{-}$. Upon substituting these into (\ref{ADMmassC}) we recover
(\ref{ADMmassA}). 

Equally we could have started with equation (\ref{ADMmassA}) and used it to 
derive equation (\ref{ADMmassC}). 

\bibliographystyle{unsrt}
\bibliography{brewin,regge,adm-mass,other,books}

\def\bold#1{\setbox0=\hbox{#1}
\kern-0.015em\copy0\kern-\wd0%
\kern+0.030em\copy0\kern-\wd0%
\kern-0.015em\raise0.0233em\copy0\kern-\wd0%
\lower0.0233em\copy0\kern-\wd0%
\raise0.0133em\copy0\kern-\wd0%
\lower0.0133em\copy0\kern-\wd0%
\box0 }
\def\Hrule{\hrule height0.5pt}%
\def\Vrule{\vrule width0.5pt}%
\def\RuledBox#1{\hbox{\Vrule\vtop{%
\vbox{\Hrule\kern0pt\hbox{\kern0pt#1\kern0pt}}%
\kern0pt\Hrule}\Vrule}}
\newdimen\boxwidth
\newdimen\boxdepth
\newdimen\boxwidthscale
\newdimen\boxdepthscale
\newdimen\boxXoffset
\newdimen\boxYoffset
\newdimen\boxtmpX
\newdimen\boxtmpY
\def\stomp#1{\setbox0=\hbox{#1}\dp0=0pt\ht0=0pt\box0}
\def\atabs(#1,#2)#3{\boxtmpX=#1\boxtmpY=#2%
                    \advance\boxtmpX\boxXoffset%
                    \advance\boxtmpY\boxYoffset%
                    \vbox to 0pt{\kern\boxtmpY%
                    \hbox to 0pt{\kern\boxtmpX\stomp{#3}\hss}\vss}\nointerlineskip}
\def\atrel(#1,#2)#3{\boxtmpX=#1\boxwidth\boxtmpY=#2\boxdepth%
                    \advance\boxtmpX\boxXoffset%
                    \advance\boxtmpY\boxYoffset%
                    \vbox to 0pt{\kern\boxtmpY%
                    \hbox to 0pt{\kern\boxtmpX\stomp{#3}\hss}\vss}\nointerlineskip}
\def\atpxl(#1px,#2px)#3{\boxtmpX=#1\boxwidthscale\boxtmpY=#2\boxdepthscale%
                        \advance\boxtmpX\boxXoffset%
                        \advance\boxtmpY\boxYoffset%
                        \vbox to 0pt{\kern\boxtmpY%
                        \hbox to 0pt{\kern\boxtmpX\stomp{#3}\hss}\vss}\nointerlineskip}
\def\DrawBox(#1cm,#2cm,#3px,#4px)#5{%
\boxwidth=#1cm%
\boxdepth=#2cm%
\boxwidthscale=\boxwidth\divide\boxwidthscale#3%
\boxdepthscale=\boxdepth\divide\boxdepthscale#4%
\hbox to \boxwidth{\vtop to \boxdepth{#5\vfill}\hfill}}
\def\SetOffset(#1cm,#2cm){\boxXoffset=#1cm\boxYoffset=#2cm}
%
\parindent=0pt
%
\SetOffset(0cm,0cm)%
{\hbox to \textwidth{\hfill{%
\DrawBox(13.5cm,18.5cm,638px,890px){%
\SetOffset(-0.75cm,-2cm)%
\atpxl(110px,540px){{\includegraphics[width=0.75\boxwidth]{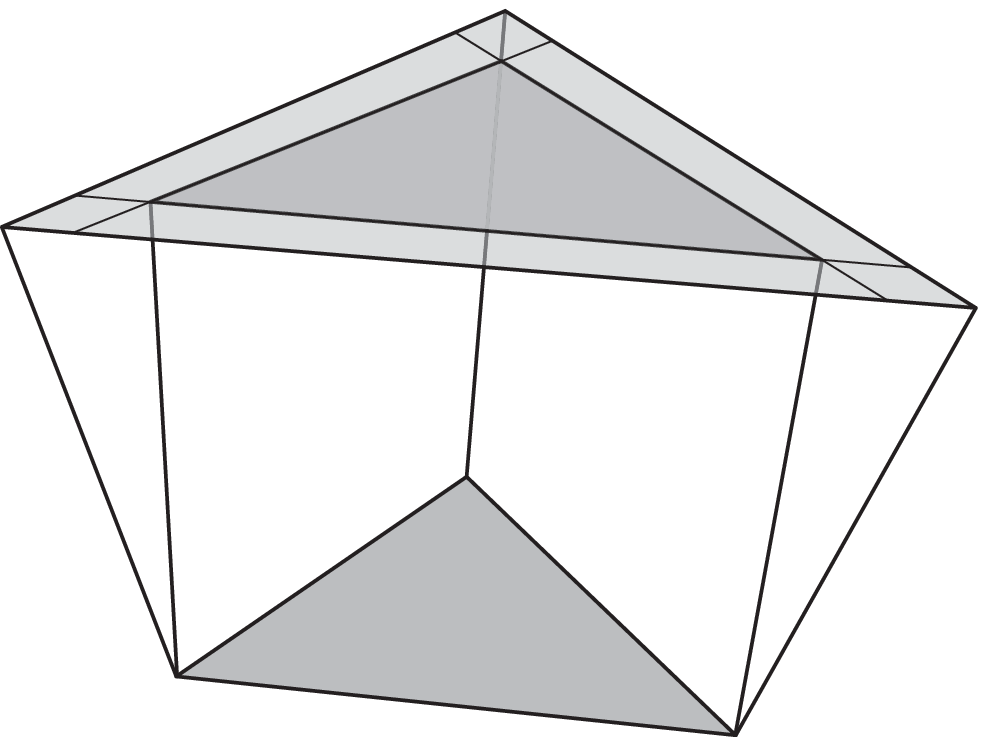}}}
\atpxl(370px,564px){\bold{\Large$L_{ab}$}}
\atpxl(431px,348px){\bold{\Large$L_{ab+}$}}
\atpxl(160px,520px){\bold{\Large$a$}}
\atpxl(496px,555px){\bold{\Large$b$}}
\atpxl(360px,407px){\bold{\Large$c$}}
\atpxl( 75px,275px){\bold{\Large$a_+$}}
\atpxl(606px,317px){\bold{\Large$b_+$}}
\atpxl(360px,158px){\bold{\Large$c_+$}}
\atpxl(245px,272px){\bold{$a'$}}
\atpxl(450px,292px){\bold{$b'$}}
\atpxl(353px,233px){\bold{$c'$}}
\atpxl(207px,403px){\bold{\Large$\delta s$}}
\atpxl(324px,491px){\bold{\Large$S_i$}}
\atpxl(339px,268px){\bold{\Large$S_{i+}$}}
\SetOffset(0cm,-0.5cm)%
\atrel(0,1.0){\hsize=\boxwidth{\hbox{\vbox{%
This image contains two cylinders each with a triangular crossection. The outer
cylinder, based on the vertices $(a,b,c)$ and $(a,b,c)_{+}$ is a cylinder of the
simplicial lattice. The inner cylinder, based on the vertices $(a,b,c)$ and
$(a,b,c)'$, is built from the simplicial lattice by dragging the lower triangle
$(a,b,c)$ outward along its normal a distance $\delta s$. The areas of the
two grey triangles are thus equal and thus the change in area from $S_i$ to
$S_{i+}$ equals the sum of the areas of the three (approximate) rectangles plus the
three diamonds on $S_{i+}$. These later areas vary as $\delta s^2$ and thus, to
leading order, the contribution to $\delta A$ from the leg $(a,b)$ will be $\delta A_{ab} = \theta_{ab}L_{ab}\delta s$ where $\theta_{ab}$ is the
angle between the faces $(a,b,a',b')$ and $(a,b,a_+,b_+)$.}}}}
}}\hfill}}
\end{document}